\documentclass[aps,prc,twocolumn,showpacs,amsmath,amssymb,nofootinbib]{revtex4}
\usepackage {lscape}
\usepackage{graphicx}
\usepackage{amsmath}
\usepackage{color}

\usepackage{epsfig,colordvi}

\usepackage{graphicx}
\usepackage{dcolumn}
\usepackage{bm}
\usepackage{epsfig}
\usepackage{array}
\usepackage{amsmath}
\usepackage[final]{pdfpages}
\usepackage{pdfpages}

\usepackage[bookmarks=true]{hyperref}
\usepackage{color}

  \hypersetup{
   colorlinks=true,
   pdfborder={0 0 0},
   citecolor=blue,
   linkcolor=blue,
   urlcolor=blue
 }

\begin{document}

\title{Ab Initio No Core Shell Model Study of Neutron Rich Nitrogen Isotopes}

\author{ Archana Saxena and Praveen C. Srivastava\footnote{Corresponding author: pcsrifph@iitr.ac.in}}
\address{Department of Physics, Indian Institute of Technology Roorkee, Roorkee
247 667, India}


\date{\hfill \today}

\begin{abstract}
In the present paper, we have calculated the energy spectra for neutron rich 
 $^{18-22}$N  isotopes using no core shell model (NCSM).
 To calculate the energy spectrum we have used three different $NN$ potentials, inside non-local outside Yukawa (INOY),
 next-to-next-to-next-leading order (N3LO) from chiral effective field theory and charge-dependent Bonn 2000 (CDB2K).
 The INOY potential, which is a two body interaction but also have the effect of three body forces by short range and non local
 character present in it. The calculations have been done at  $\hbar\Omega$=20 MeV, 14 MeV and 12 MeV using INOY, N3LO and CDB2K
 potentials, respectively. Apart from this, we have also performed  shell model calculations with the  YSOX interaction.
 The results with INOY interaction show good agreement with the experimental data in comparison to other three interactions. We have also shown 
 the occupancy of different orbitals involved corresponding to the largest model space ($N_{max}$= 4) in the present calculations.
\end{abstract}

\pacs{21.60.Cs, 21.30.Fe, 21.10.Dr, 27.20.+n, 27.30.+t}
\maketitle

\vspace{1cm}

\section{\bf{Introduction}}
In nuclear physics, solving many body problem from first principle is computationally hard. But now a days, an advancement in computational
facility made it possible. There are many $ab~initio$ methods available to study nuclear properties. The no core shell model 
\cite{Navratil3,Barrett,Navratil1,Navratil2}  is one of them. At present NCSM is well established technique used in nuclear physics to
calculate nuclear properties. Here, we solve $A$-body Schr\"{o}dinger equation for the particles treated as non relativistically and interacted by 
realistic two  body forces. With the NCSM,
a detailed study has been done for even carbon isotopes where ground state energy, quadruple moment of ${2^{+}_{1}}$ state, 
some $B(E2)$ transitions and occupancies of ${0^{+}_{1}}$ and ${2^{+}_{1}}$ are calculated \cite{Carbon_isotopes} 
using INOY \cite {Doleschall_1,Doleschall_2} and CDB2K \cite{CDB2K} interactions. 

In the present work we will study the nitrogen  isotopes and mainly focused on neutron rich side. 
The structure of neutron rich nuclei $^{19-22}$N has been studied by in-beam $\gamma$-ray spectroscopy and spectra and other properties are
compared with shell model calculations using WBT and WBTM interactions, where $N=14$ closed sub shell is discussed \cite{Nitrogen}.
The $^{22}$N has halo structure in its ground state \cite{22N1,22N2}.
Recently, the point proton radii of neutron rich $^{17-22}$N isotopes have been measured from charge changing cross section in Ref. \cite{proton_radii}.
More recently, Yuan and Suzuki $\it{et~al}$, have done systematic study of B to O isotopes with a interaction YSOX 
 which include (0-3) $\hbar \Omega$ excitations \cite{YSOX} in full $psd$ model space. To the best of 
 our knowledge for the first time we have done systematic NCSM calculations for nitrogen isotopes. 


The present paper is organized as follows: In Sec. \ref{Form}, the theory and formalism of NCSM is given, In Secs. \ref{int} and \ref{details},
we have  discussed  about effective interactions which are used in calculations and details of the calculations, respectively. 
The results and discussions part is in Sec. \ref{Results} and in the end we conclude the paper in Sec. \ref{conclusion}.

\section{\bf{No Core Shell Model Formalism} \label{Form}}

The starting A-body Hamiltonian is given by:
\begin{eqnarray}
\label{eq:(1)}
H_{A} = T_{rel} + V  = \frac{1}{A} \sum_{i< j}^{A} \frac{({\vec p_i - \vec p_j})^2}{2m} 
 +\sum_{i<j}^A V_{NN,ij} 
 \label{bareham}
\end{eqnarray}

$T_{rel}$ is the relative kinetic energy.  The momenta of the individual nucleons are given by $p_{i}$ (i=1,.....A).
The nucleon mass is given by $m$.
In the present work we have dealt with the two body part only. The $V_{NN,ij}$ is the \textit{NN} interaction having nuclear and Coulomb part both.  
Next, we divide the $A$-nucleon infinite HO basis space into finite active space ($P$) having all states of up to $N_{max}$ HO excitations above
the unperturbed ground state and an
excluded space ($Q=1-P$).

\begin{figure}
\begin{center}
\includegraphics[width=10cm,height=8cm,clip]{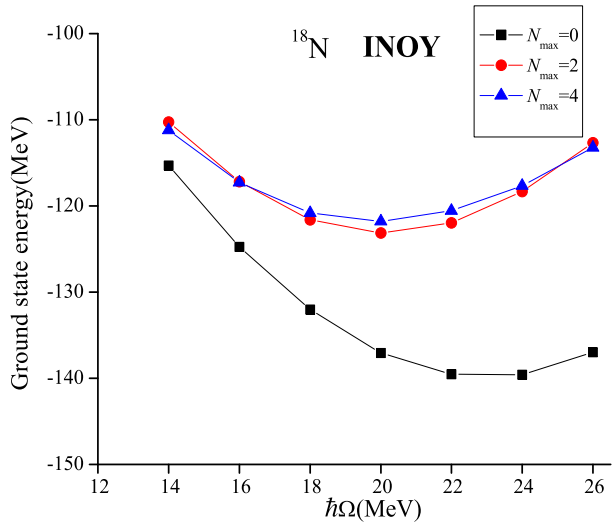}
\caption{\label{hw_curve}The variation of g. s. energy with different frequencies and different model space sizes.}
\end{center}  
\end{figure}

\begin{figure*}
\begin{center}
\includegraphics[width=15.8cm,height=7.5cm,clip]{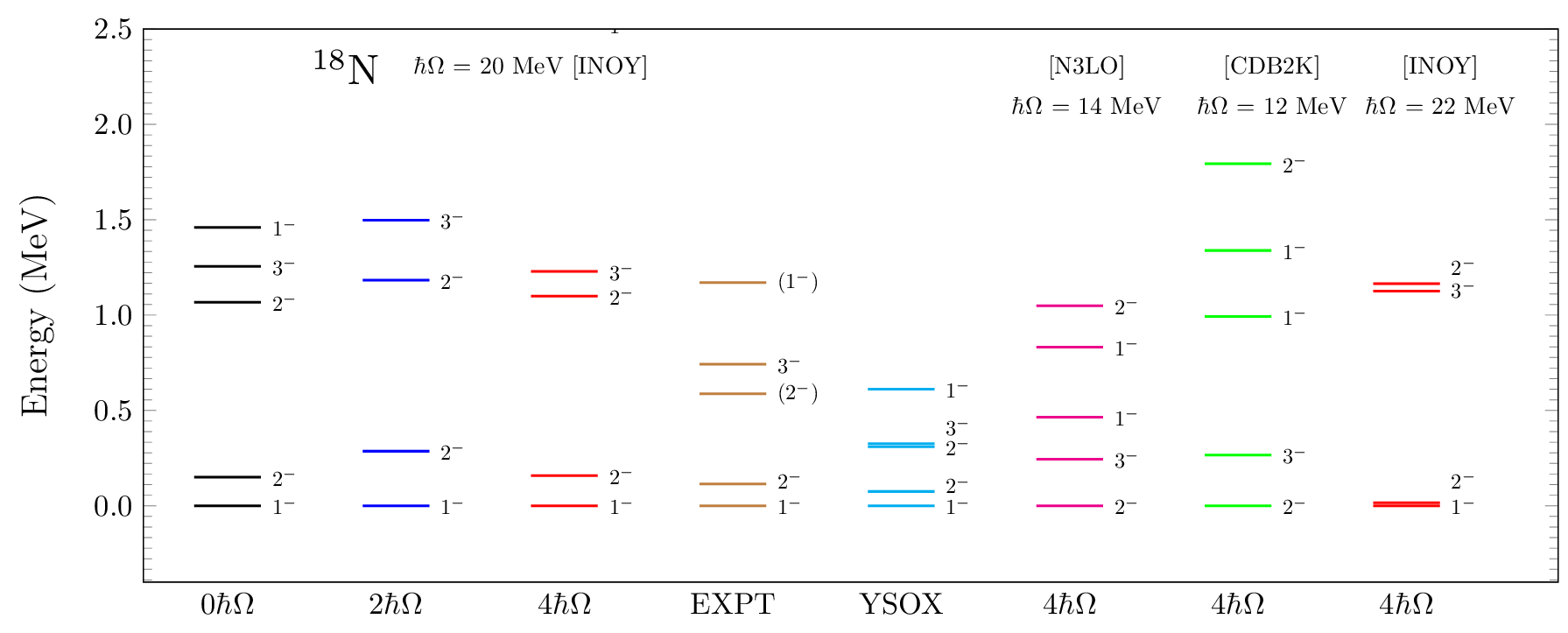}
\includegraphics[width=15.8cm,height=7.5cm,clip]{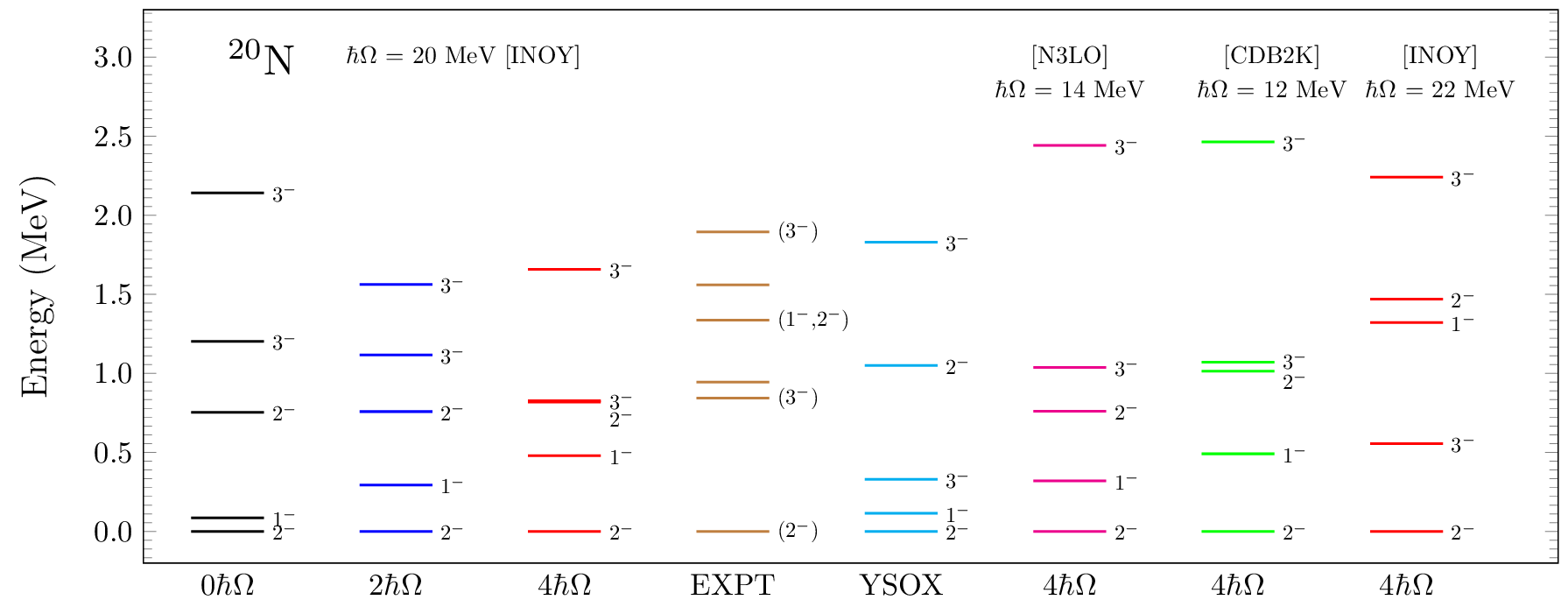}
\includegraphics[width=15.8cm,height=7.5cm,clip]{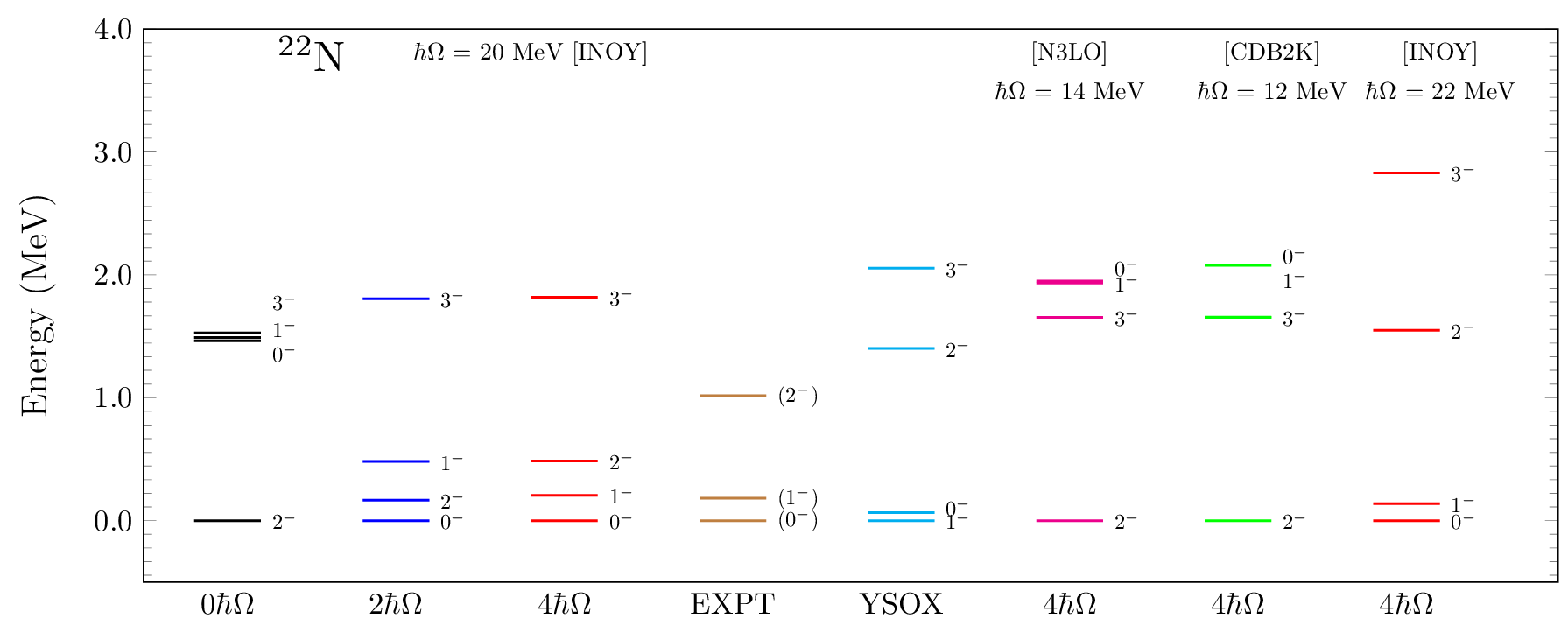}
\caption{\label{N1_spectra}The energy spectra of $^{18,20,22}$N with different model space sizes. The experimental data is taken from Refs. \cite{nndc,18N}.}
\end{center}
\end{figure*}

\begin{figure*}
\begin{center}
\includegraphics[width=15.8cm,height=7.5cm,clip]{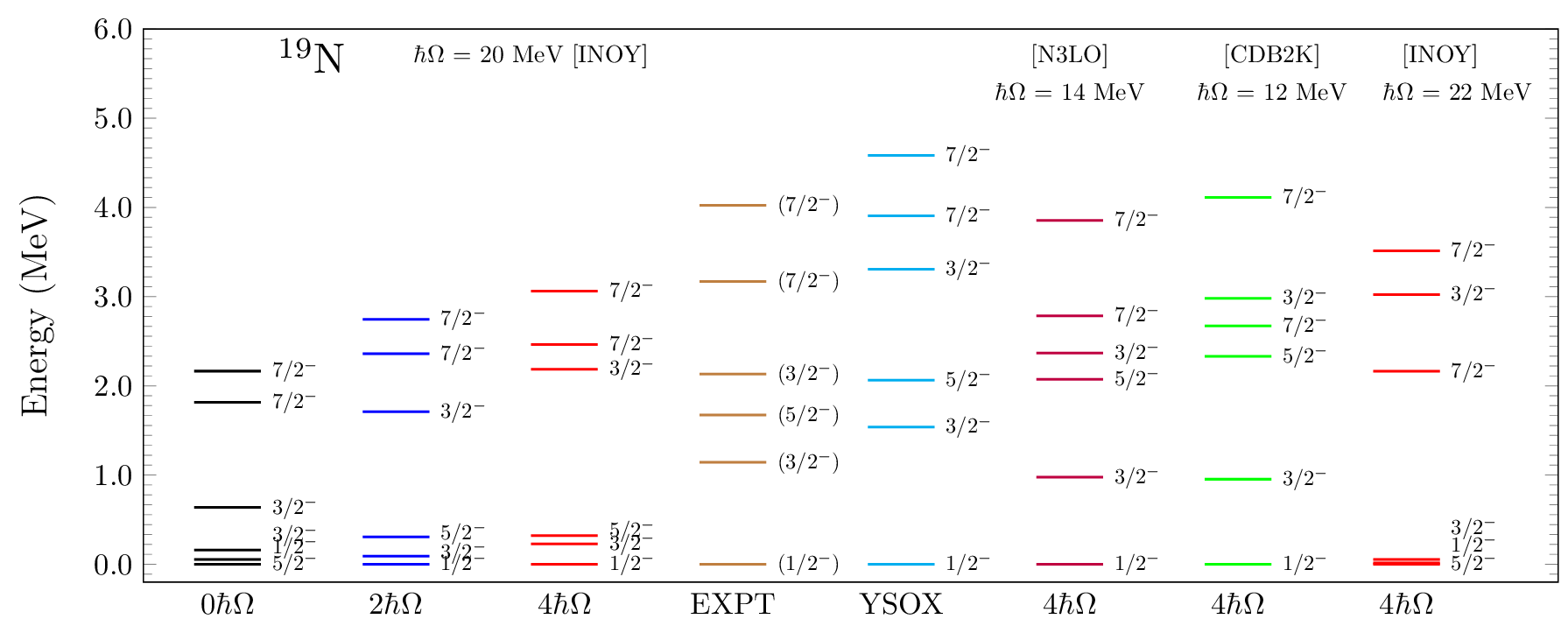}
\includegraphics[width=15.8cm,height=7.5cm,clip]{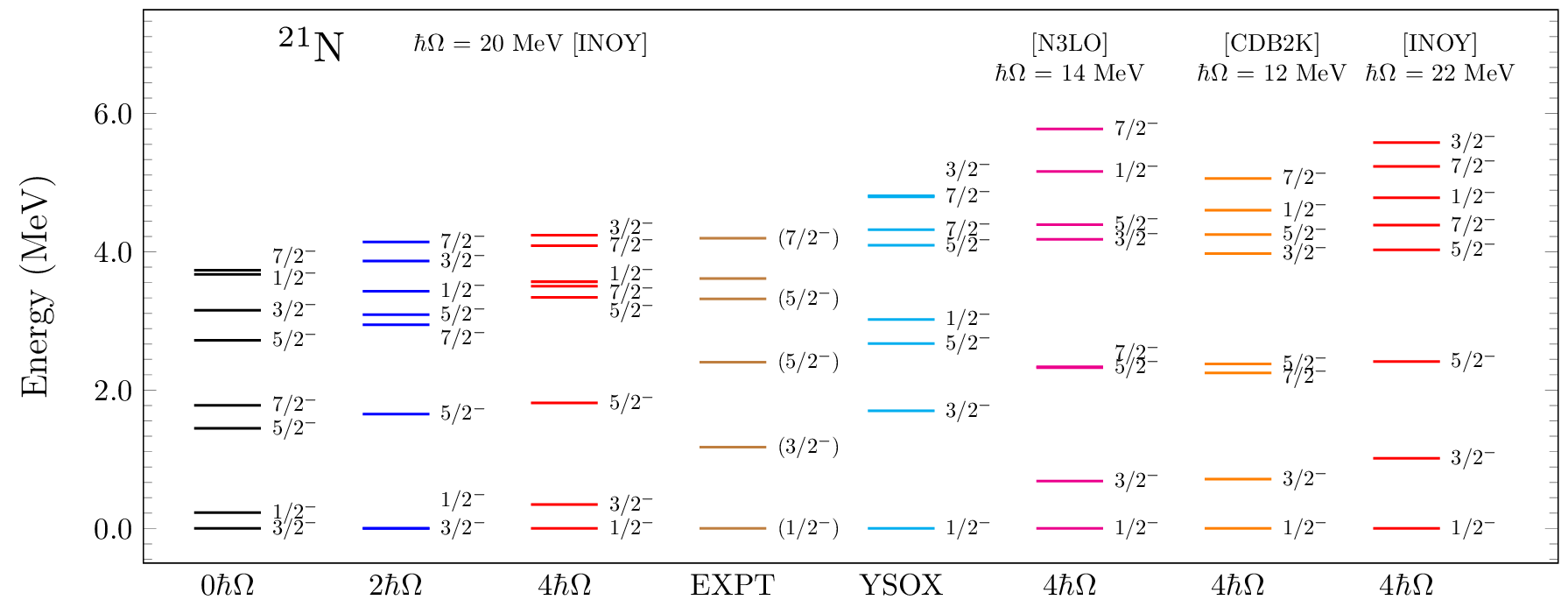}
\caption{\label{N2_spectra}The energy spectra of $^{19,21}$N with different model space sizes. The experimental data is taken from Ref. \cite{nndc}.}
\end{center}
\end{figure*}

\begin{figure}
\begin{center}
\includegraphics[width=8.8cm,height=7cm,clip]{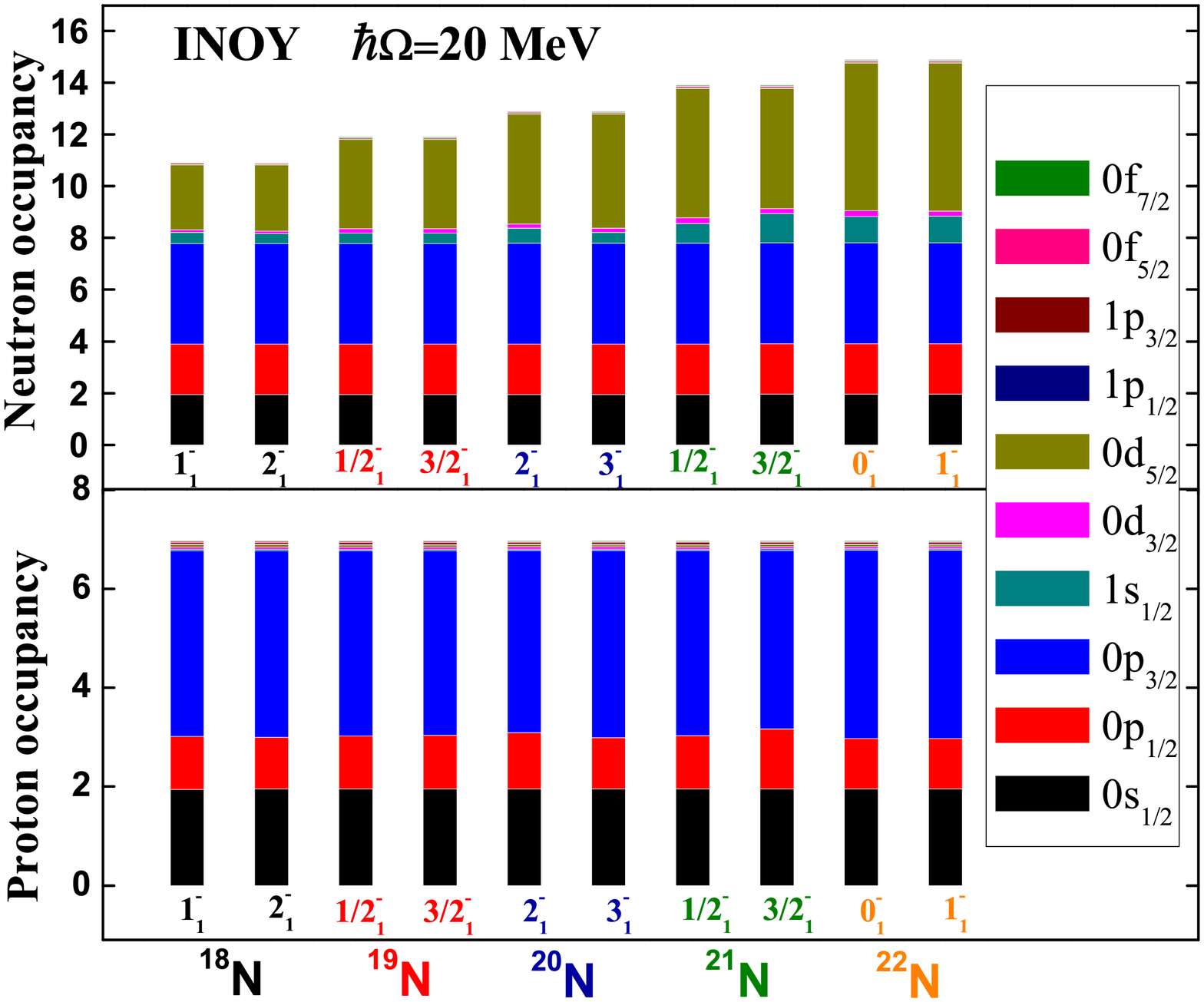}
\includegraphics[width=8.8cm,height=7cm,clip]{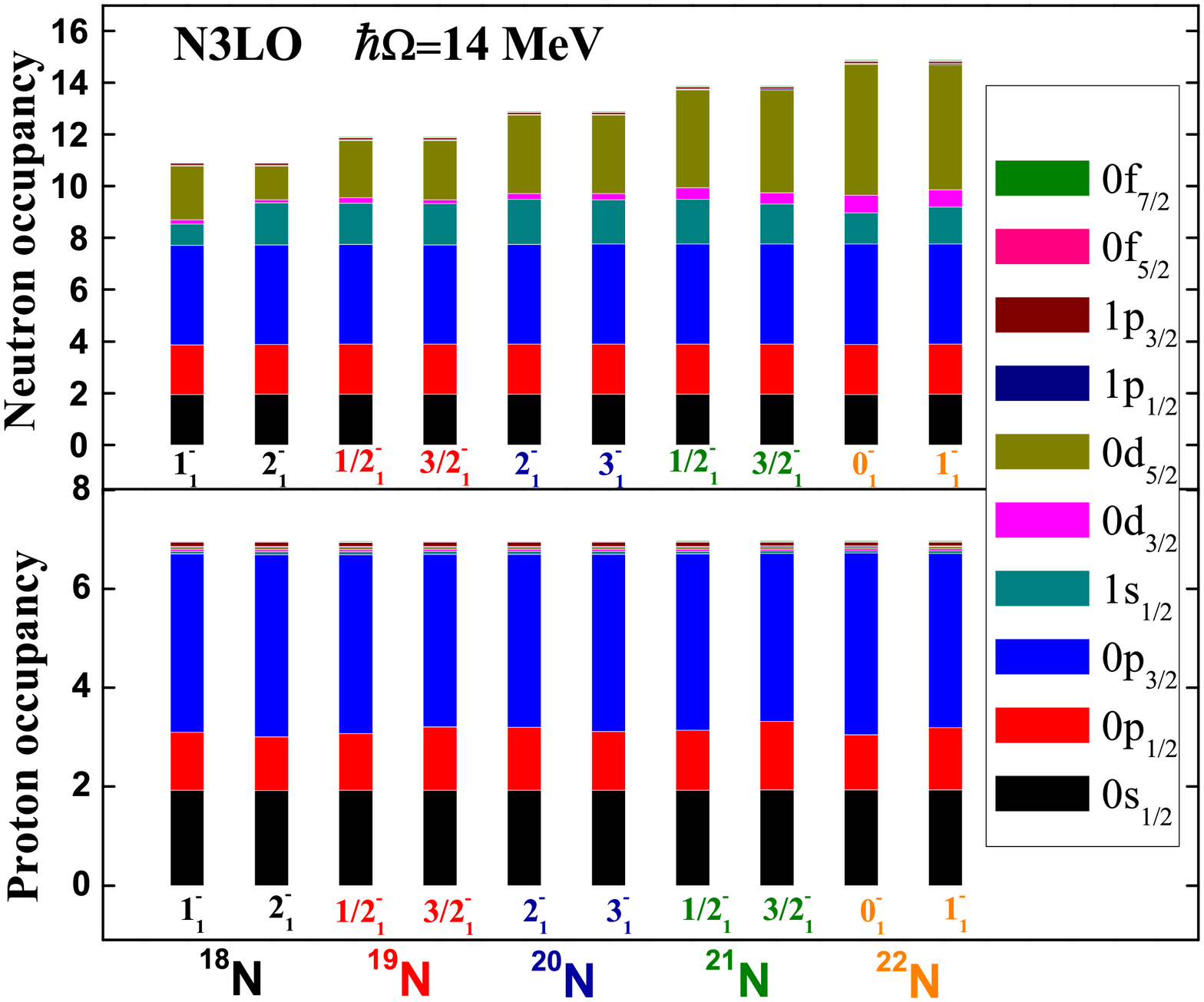}
\includegraphics[width=8.8cm,height=7cm,clip]{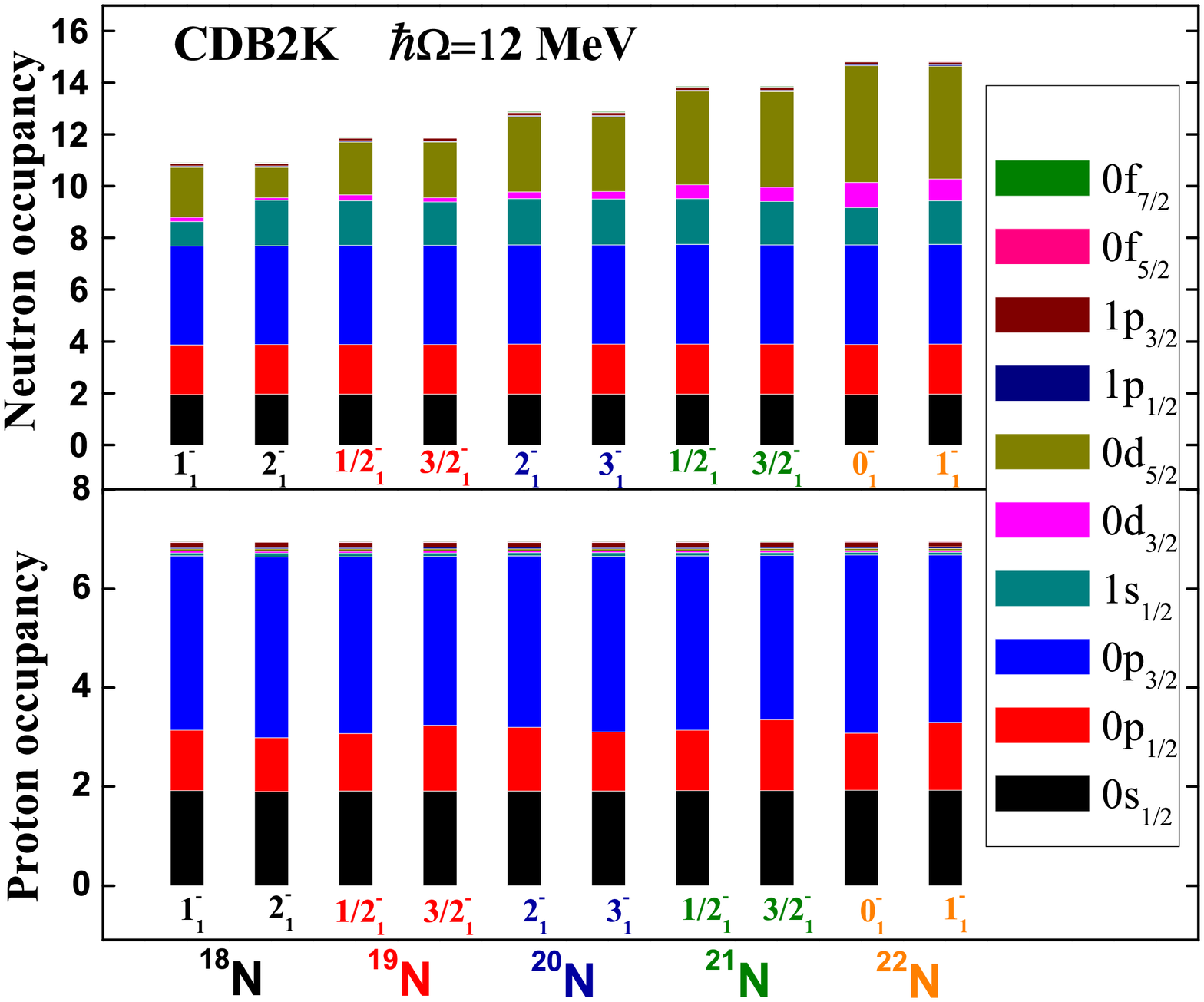}
\caption{\label{occu_N}The occupancy of different orbits for nitrogen isotopes using INOY, N3LO and CDB2K interaction.}
\end{center}
\end{figure}

\begin{figure}
\begin{center}
\includegraphics[width=8.8cm,height=7.5cm,clip]{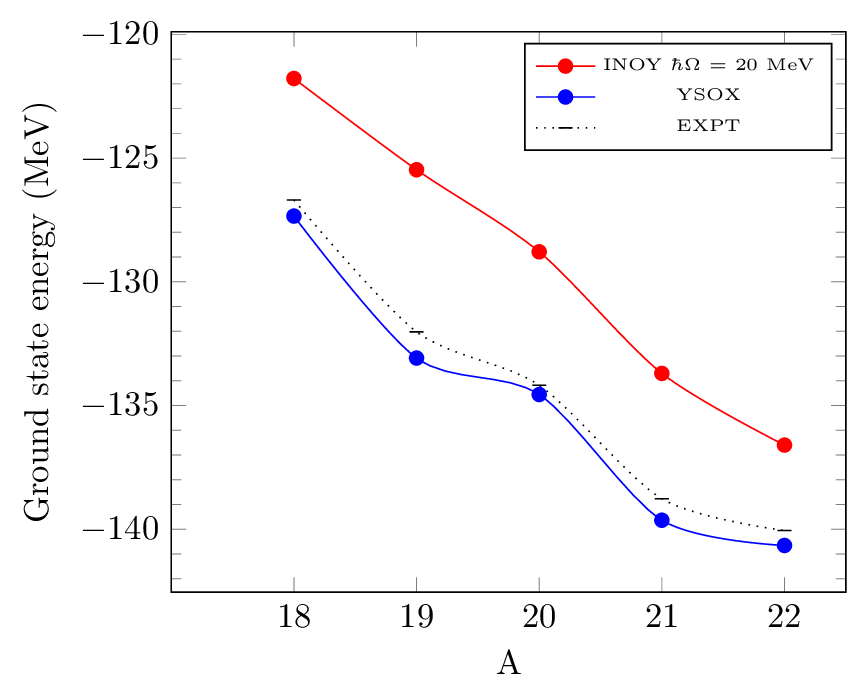}
\caption{\label{gs} Comparison of calculated and experimental g.s. energies of N isotopes with INOY and YSOX interactions.}
\end{center}
\end{figure}

We add the center-of-mass (c.m.) harmonic oscillator (HO) Hamiltonian $H_{c.m.}$($= T_{c.m.}+ U_{c.m.}$, $U_{c.m.}=\frac{1}{2}Am{\Omega}^2 {\vec R}^2$, $\vec R=(\frac{1}{A})\sum_{i=1}^A {\vec r_{i}}$.) to Eq. \ref{bareham}. As we use slater determinant basis, 
the Lawson Projection term \cite{lawson} is added to shift the spurious states (arises from the incorrect treatment of the centre-of-mass motion) 
to the Eq. \ref{bareham}.
The Hamiltonian used in final calculations is given by:
  \begin{eqnarray}
H_{A,eff}^{\Omega} = P \Bigg\{ \sum_{i=1}^A \Bigg[ \frac{({\vec p_i - \vec p_j})^2}{2mA} + \frac{m {\Omega}^2 }{2A}({\vec r_i-\vec r_j})^2  
 +V_{ij,eff}^{\Omega,A} \Bigg] \nonumber\\
+ \beta \left( H_{c.m.} - \frac{3}{2} \hbar \Omega \right)  \Bigg\} P. 
\label{Finalham}
\end{eqnarray} 

Where $\beta$ is a parameter which is equal to 10.0 in the present calculations.
The Eq. \ref{Finalham} is a Hamiltonian which we get after applying unitary transformation because we are not using soft interactions (to soften the potential with the purpose 
of simplifying many-body calculations, 
these interactions are obtained by applying the
unitary transformation to the two-nucleon system in momentum space with a regulator).  So, we need a renormalization 
scheme to soften the interactions. Here, we use  Okubo-Lee-Suzuki (OLS) scheme \cite{OLS1, Suzuki1, Suzuki2}. Now, we get an effective Hamiltonian which is in $A$- body space.
In our calculations we have done NCSM calculations with the renormalized interactions keeping up to two-body cluster terms.
 
In the present paper, for the NCSM calculations, we have used the pAntoine \cite{Caurier1, CFors1} shell model code which is adapted
to NCSM \cite {Caurier3}. 
In the case of $^{22}$N, for the largest model space $N_{max}$= 4, the corresponding dimension is $\sim 6.4\times 10^{7}$. 
We have compared the NCSM results with the shell model calculations using  YSOX interaction. For shell model calculations we have used 
KSHELL code \cite{kshell}.

\section{\bf{Effective $NN$ Interaction} \label{int}}
In the present work we have studied the neutron rich nitrogen isotopes with the three different $NN$ interactions: 
INOY, CDB2K and N3LO \cite{Entem, Machleidt2, Rup}. 
 The magnitude of higher body forces decreases as we go from two body to higher body but still they are important to study some properties of
 nuclei for e.g. the drip-line in oxygen isotopes can be explained only with the inclusion of three body forces \cite{Odrip}. In the
INOY potential, a non-local potential in coordinate space, is a mixture of local and non-local parts. The behaviour of INOY is local Yukawa
tail at longer ranges ($>$ 3 fm) and non-local at short range. 
The form of INOY $NN$ interaction is given in Refs. \cite{Doleschall_1,Doleschall_2}. This interaction reproduces the $^{3}$H and $^{3}$He binding energy accurately and results are in agreement with the experimental data without adding 3N force.
The CDB2K interaction is also nonlocal interaction and charge dependent. The charge dependency is introduced due to pion mass splitting. 
This potential fits the p-p data below 350 MeV which was available in the year 2000.
The N3LO interaction is from chiral effective field theory. Here, we use only $NN$ part. 


\section{\bf {Details of the Calculations} \label{details}}
In the present work we perform calculations for nitrogen isotopes. As we know NCSM calculations are variational,
depend on HO frequency $\hbar\Omega$ and size of the model space $N_{max}$. To see this dependence, we have calculated the g.s. energy with different $N_{max}$ and $\hbar\Omega$, see Fig. \ref{hw_curve}. We are interested to see that region in which the dependence of g.s. energy on frequency is minimum (for largest model space).
We select that frequency for our NCSM calculations. This procedure is called optimization of frequency.
When we use this frequency, we get faster convergence ( computational time will be smaller) rather than other values of frequencies. This is the benefit for doing optimization of frequency.
So, we have done our calculations with frequency $\hbar\Omega$= 20 MeV. 
 For the other interactions we have chosen the frequency from the literature which is suitable in this mass 
region. 
We have chosen the frequency $\hbar\Omega$=20 MeV for INOY and  $\hbar\Omega$=14 MeV for N3LO interaction \cite{Barrett}.
In the case of CDB2K, we have taken $\hbar\Omega$=12 MeV \cite{Carbon_isotopes}.

\section{\bf {Results and discussions} \label{Results}}

We have done calculations using INOY at $\hbar\Omega$= 20 MeV, CDB2K and N3LO interactions at 12 and 14 MeV, respectively. 
We have also compared our INOY results at $\hbar\Omega$= 22 MeV.  
The energy spectra are  shown in Figs. \ref{N1_spectra} and \ref{N2_spectra}.
In the case of $^{18}$N, the g.s. is correctly reproduced by INOY $NN$ and  YSOX interaction, while other two 
interactions give $2^{-}$ as a g.s. The order of energy states are correct with the INOY ($\hbar\Omega$=20 MeV) and YSOX only. The calculated $1_2^-$ state is at higher energy ($>$ 2.5 MeV) with INOY interaction (except for $N_{max}=0$).
The NCSM results for $N_{max}= 4$ with INOY ($\hbar\Omega$=22 MeV) are compressed in comparison to the CDB2K interaction.

For $^{20}$N, the results with the INOY ($\hbar\Omega$=22 MeV) interaction are better than other interactions. Although the g.s. is correctly 
reproduced by all the three interactions but the higher states are not in agreement with the N3LO and CDB2K interactions. The first $3^{-}$ state is close to the experimental data with INOY ($\hbar\Omega$=20 MeV) and $1^{-}$ is close to experimental data with INOY ($\hbar\Omega$=22 MeV).

In the case of $^{22}$N, only INOY interaction can reproduce the correct g.s. $0^{-}$ and level ordering with both the frequencies.
All the other interactions are not able to produce correct g.s. and level ordering of the energy states.

In the case of $^{19}$N, INOY ($\hbar\Omega$=20 MeV) and the other interactions reproduce the correct g.s. $1/2^{-}$,
though, all the states are not yet been confirmed experimentally. 
The g.s. and first two excited states are very compressed with the INOY at both the frequencies in comparison to the other interactions.
The N3LO interaction gives the energy states better and level ordering is correct with the experimental one. Overall the INOY interaction gives compressed energy levels.

For $^{21}$N, the g.s. is correctly reproduced. Higher states are not yet been confirmed experimentally. 
All the interactions give first excited state as $3/2^{-}$. Similarly, the second excited state seems to be 
$5/2^{-}$. For higher states, we are not sure for spin prediction.
So, from our NCSM calculations it is clear that INOY interaction which has the effect of three body forces is suitable to study the  neutron rich nitrogen isotopes.
The inclusion of  3N forces is important to reproduce correct spectra with CDB2K and N3LO interactions.

In Fig. \ref{occu_N}, we have shown the occupancy of first two states of nitrogen isotopes with the INOY ($\hbar\Omega$=20 MeV), CDB2K, 
and N3LO interactions correspond to $N_{max}$= 4 model space size. For $N_{max}$= 4, we have taken 28 orbitals.
Here, we have shown the occupancy up to $fp$ space because the occupancy of higher orbitals are very small to visualize. 
Although, the magnitudes of  occupancies of higher orbitals are very small, still they are important in the calculation. 
The contribution of neutron occupancy from 0$d_{3/2}$ and 1$s_{1/2}$ orbitals for CDB2K and N3LO 
interaction is larger in comparison to INOY interaction. This larger occupancy is also reflected in 
the energy spectra. The CDB2K and N3LO results are similar for the g.s. spin and first excited state, however
the occupancies for INOY interaction is different and for this interaction we are getting 
results which differ from other two interactions.
In Fig. \ref{gs}, the calculated g.s. energy for $^{18-22}$N isotopes using INOY and YSOX interactions follow
the same trend as the experimental data. The g.s. energy for nitrogen isotopes with the other interactions are 
given in the Table \ref{GS_table} in which results with N3LO and CDB2K are very far from the experimental data.
If we go to higher $N_{max}$, the results will come closer to the experimental g.s. energies.

\begin{table}
\centering 
\caption{The g.s. energies (in MeV) for nitrogen isotopes using YSOX, INOY ($\hbar\Omega$=20 MeV), N3LO ($\hbar\Omega$=14 MeV), 
and CDB2K ($\hbar\Omega$=12 MeV) interactions.}
\label{GS_table}
\vspace{0.7cm}
\begin{tabular}{|l|l|l|l|l|l|}
\hline
Nucleus & EXP & YSOX & INOY  & N3LO & CDB2K\\
\hline
$^{18}$N & -126.695  & -127.344  & -121.782 & -112.036 & -102.979 \\
\hline
$^{19}$N & -132.025 & -133.083  & -125.471 & -117.084  & -107.616 \\
\hline
$^{20}$N & -134.180 & -134.556   & -128.788  & -119.857  & -109.921 \\
\hline
$^{21}$N & -138.768 & -139.637  & -133.702   & -124.769  & -114.278 \\
\hline
$^{22}$N & -140.052 &  -140.657 & -136.560  & -127.114  & -116.052 \\
\hline
\end{tabular}
\end{table}


\section{\bf{Conclusions} \label{conclusion}} 

In the present work, we have performed NCSM calculations with different interactions (INOY, N3LO and CDB2K) for neutron rich nitrogen isotopes.
We have also compared our NCSM results with recently developed YSOX interaction for $psd$ space from the Tokyo group.
  In $^{18}$N, the INOY and  YSOX interaction predict second excited state as $2^-$.
  For $^{20}$N, the results of INOY ($\hbar\Omega$=22 MeV) interaction are better than  YSOX interaction.
  For $^{22}$N, the INOY results for ground and first excited states are better than  YSOX interaction.
The N3LO and CDB2K interactions are unable to predict correct ground state.
 For $^{19}$N, the NCSM results with N3LO are much better.

\section*{Acknowledgement:}
 AS acknowledges financial support from MHRD (Govt. of India) for her Ph.D. thesis work.
We would like to thank Prof. Petr Navr\'{a}til  for providing us his NN effective interaction code and
Prof. Christian Forss\'{e}n for pAntoine.
We would also like to thank Prof. Ruprecht Machleidt for valuable comments on this article
and Prof. Toshio Suzuki for the YSOX interaction.
PCS acknowledges the hospitality extended to him during his stay at TRIUMF.



\begin{thebibliography}{10}
\providecommand{\url}[1]{\texttt{#1}}
\providecommand{\urlprefix}{URL}
\providecommand{\eprint}[2][]{\url{#2}}




\bibitem{Navratil3}
P. Navr\'{a}til, S. Quaglioni,  I. Stetcu and B.R. Barrett, Recent developments in no-core shell-model calculations, {\color{blue} J. Phys. G: Nucl. Part. Phys.  \textbf{36}, 083101 (2009).}



\bibitem{Barrett}
Bruce R. Barrett, Petr Navr\'{a}til and James P. Vary, {\color{blue} Progress in Particle and Nuclear Physics \textbf{69}, 131-181 (2013).}

\bibitem{Navratil1}
P. Navr\'{a}til, J.P. Vary, B.R. Barrett, Properties of ${}^{12}$C in the Ab Initio Nuclear Shell Model, {\color{blue} Phys. Rev. Lett. \textbf {84}, 5728 (2000).}

\bibitem{Navratil2}
P. Navr\'{a}til, J.P. Vary, B.R. Barrett, Large-basis ab initio no-core shell model and its application to ${}^{12}\mathbf{C}$, {\color{blue}  Phys. Rev. C \textbf {62}, 054311 (2000).}





\bibitem{Carbon_isotopes}
C Forss\'{e}n, R Roth and P Navr\'{a}til, Systematics of $2^{+}$  states in C isotopes from the no-core shell model, {\color{blue} 
J. Phys. G: Nucl. Part. Phys. \textbf{40} 055105 (2013).}

\bibitem{Doleschall_1}
P. Doleschall, and I. Borb\'ely, Properties of the nonlocal $\mathrm{NN}$ interactions required for the correct triton binding energy, {\color{blue} 
Phys. Rev. C \textbf{62}, 054004 (2000).}

\bibitem{Doleschall_2}
P. Doleschall, I. Borb\'ely,  Z. Papp, and W. Plessas, Nonlocality in the nucleon-nucleon interaction and three-nucleon bound states, 
{\color{blue} Phys. Rev. C \textbf{67}, 064005 (2003).}

\bibitem{CDB2K}
R. Machleidt, High-precision, charge-dependent Bonn nucleon-nucleon potential, {\color{blue} Phys. Rev. C \textbf{63}, 024001 (2001)}.

\bibitem{Nitrogen}

D. Sohler, M. Stanoiu, Zs. Dombr\'{a}di, F. Azaiez, B. A. Brown, M. G. Saint-Laurent, O. Sorlin, Yu.-E. Penionzhkevich,
N. L. Achouri, and J. C. Ang\'{e}lique $et. al.$, In-beam \ensuremath{\gamma}-ray spectroscopy of the neutron-rich nitrogen isotopes $^{19\ensuremath{-}22}\mathrm{N}$, {\color{blue}  Phys. Rev. C \textbf{77}, 044303 (2008).}

\bibitem{22N1}
C. S. Sumithrarachchi, D. J. Morrissey, A. D. Davies, D. A. Davies,
M. Facina, E. Kwan, P. F. Mantica, M. Portillo, Y. Shimbara, J. Stoker $et.al.$, States in $^{22}\mathrm{O}$ via $\ensuremath{\beta}$ decay of $^{22}\mathrm{N}$, {\color{blue} Phys. Rev. C \textbf{81}, 014302 (2010).}

\bibitem{22N2}
 C. Rodr\'{i}guez-Tajes, D. Cortina-Gil, H. \'{A}lvarez-Pol, T. Aumann, E. Benjamim, J. Benlliure, M. J. G. Borge,
M. Caama\~{n}o, E. Casarejos, A. Chatillon $et. al.,$ Structure of $^{22}\mathrm{N}$ and the $N=14$ subshell, {\color{blue} Phys. Rev. C \textbf{83}, 064313 (2011)}.


\bibitem{proton_radii}
S. Bagchi, R. Kanungo, W. Horiuchi, G. Hagen, T.D. Morris, S.R. Stroberg, T. Suzuki, F. Ameil, J. Atkinson,Y. Ayyad 
$et. al.,$ Neutron skin and signature of the N = 14 shell gap found from measured proton radii of $^{17-22}$N, {\color{blue} Phys. Lett. B \textbf{790}, 251  (2019).}



\bibitem{YSOX}
C. Yuan, T. Suzuki,  T. Otsuka,  F. Xu, and N. Tsunoda, Shell-model study of boron, carbon, nitrogen, and oxygen isotopes with a monopole-based universal interaction, {\color{blue}  Phys. Rev. C \textbf {85}, 064324 (2012)}.







\bibitem{lawson}
 D.H. Gloeckner and R.D. Lawson, Spurious center-of-mass motion, Phys. Lett. B \textbf{53}, 313 (1974).

\bibitem{Suzuki1} 
K. Suzuki and S. Y. Lee, Convergent Theory for Effective Interaction in Nuclei, {\color{blue} Prog. Theor. Phys. \textbf{64}, 2091 (1980).}

\bibitem{Suzuki2}  
K. Suzuki, Construction of Hermitian Effective Interaction in Nuclei— General Relation between Hermitian and Non-Hermitian Forms, {\color{blue} Prog. Theor. Phys. \textbf{68} (1), 246 (1982).}



\bibitem{OLS1}
S. Okubo, Diagonalization of Hamiltonian and Tamm-Dancoff Equation, {\color{blue} Progr. Theor. Phys.  \textbf{12},  603 (1954).}





\bibitem{Caurier1}
E. Caurier and F. Nowacki, Present status of shell model techniques, {\color{blue} Acta Phys. Pol. B \textbf {30} (3), 705 (1999).}


\bibitem{CFors1}
C. Forss\'{e}n, B. D. Carlsson, H. T. Johansson, and D. S\"{o}\"{o}f, Large-scale exact diagonalizations reveal low-momentum
scales of nuclei, {\color{blue} Phys. Rev. C \textbf {97}, 034328 (2018)}.


\bibitem{Caurier3}
E. Caurier, P. Navr\'{a}til, W. E. Ormand, and J. P. Vary, Intruder states in ${}^{8}\mathrm{Be}$, {\color{blue} Phys. Rev. C \textbf {64}, 051301(R) (2001).}

\bibitem{kshell}
N. Shimizu, Nuclear shell-model code for massive parallel computation, ``KSHELL", {\color{blue} arXiv:1310.5431v1 [nucl-th]}.

\bibitem {Entem}
D. R. Entem, and R. Machleidt, Accurate charge-dependent nucleon-nucleon potential at fourth order of chiral perturbation theory, {\color{blue}  Phys. Rev. C \textbf{68}, 041001(R) (2003).}


\bibitem{Machleidt2}
R. Machleidt, D.R. Entem, Chiral effective field theory and nuclear forces, {\color{blue} Phys. Rep. 503  1 (2011).}


 
 \bibitem{Rup}
 Ruprecht Machleidt (private communication).

\bibitem{Odrip}
T. Otsuka, T. Suzuki, J.D. Holt, A. Schwenk,and Y. Akaishi, Three-Body Forces and the Limit of Oxygen Isotopes, 
{\color{blue} Phys. Rev. Lett. \textbf {105}, 032501 (2010).}

\bibitem{nndc}
NNDC, https://www.nndc.bnl.gov/


\bibitem{18N}
 C. R. Hoffman, M. Albers,   M. Alcorta, S. Almaraz-Calderon,  B. B. Back,  S. I. Baker,  S. Bedoor,  P. F. Bertone,  B. P. Kay,  J. C. Lighthall, T. Palchan,  $et. al.,$ Single-neutron excitations in ${}^{18}$N, {\color{blue} Phys. Rev. C \textbf{88} 044317 (2013).}

\end{thebibliography}
\end{document}